\documentclass[12pt]{elsarticle}

\IfFileExists{srcltx.sty}{\usepackage[active]{srcltx}}

\usepackage{graphicx}

\usepackage{amssymb}

\usepackage{lineno}

\usepackage{url}

\biboptions{sort&compress}

\newcommand\araa{ARA\&A}
\newcommand\apjl{{ApJ}}

\journal{Astroparticle Physics}

\begin{document}

\begin{frontmatter}


\title{Understanding the spectrum of a distant blazar PKS~1424+240 and its implications. }


\author[ucla]{Warren Essey}
\author[ucla,ipmu]{Alexander Kusenko}

\address[ucla]{Department of Physics and Astronomy, University of California, Los Angeles, CA 90095-1547, USA}
\address[ipmu]{Kavli IPMU (WPI), University of Tokyo, Kashiwa, Chiba 277-8568, Japan}

\begin{abstract}
We calculate the gamma-ray spectrum of a distant blazar, PKS~1424+240, located at redshift $z \ge 0.6035$, where the optical depth to primary gamma rays of the highest observed energies is  $\tau\ge 5$. The spectral shape agrees well with what is expected from secondary gamma rays produced in line-of-sight interactions of cosmic rays with background photons.  
In particular, we find an acceptable agreement with the source redshifts in the range $0.6 \le z \le 1.3$.  
We discuss the implications of existing and future data for models of extragalactic background light (EBL) and for the redshift of the source.  
In the case of a future detection of temporal variability at lower energies, one can set more restrictive limits on both the source redshift and the EBL spectrum. 

\end{abstract}

\begin{keyword}
gamma rays \sep blazars \sep cosmic rays


\end{keyword}

\end{frontmatter}



\section{Introduction}

Distant blazars are believed to be powered by supermassive black holes with jets capable of accelerating electrons, protons, and nuclei to very high energies.  The gamma-ray radiation emanating from the  blazars is observed by gamma-ray telescopes.  At the higher end of the spectrum, gamma rays must interact with extragalactic background light (EBL) and lose energy through pair production.  However, line-of-sight interactions of cosmic rays with cosmic microwave background radiation (CMBR) and EBL can generate secondary gamma rays relatively close to the observer, which provides a plausible explanation of the observed hard  spectra~\cite{Essey:2009zg,Essey:2009ju,Essey:2010er,Essey:2010nd,Essey:2011wv,Murase:2011cy,Razzaque:2011jc,Prosekin:2012ne,Aharonian:2012fu,Zheng:2013lza,Kalashev:2013vba,Takami:2013gfa,Inoue:2013vpa}.  Primary gamma rays dominate the observed radiation from nearby blazars, and only the more distant sources provide an opportunity to study secondary gamma rays.  If secondary gamma rays do not contribute (for 
example, due to some large values of  intergalactic magnetic fields, or because cosmic rays are not accelerated in active galactic nuclei to the necessary energies of $10^{17}$~eV), one can consider some alternative explanations.   For example, some special structures present in the sources~\cite{Aharonian:2008su} can help
reconcile the data with theoretical predictions. The highest energy gamma rays observed from relatively nearby sources can come from secondary production in purely electromagnetic cascades~\cite{Aharonian:2001cp}.  Lorentz invariance violation or hypothetical new particles have also been invoked 
to explain the data~\cite{ex1,ex2,ex3,ex4,Simet:2007sa,Horns:2012kw,Meyer:2013fia}.  Furthermore, one may question the accuracy of redshift measurements for a number of distant blazars. 

A recent reliable measurement of PKS~1424+240 redshift $z \ge 0.6035$, combined with the data from VERITAS~\cite{Acciari:2010,Williams} and MAGIC\cite{Aleksic:2014tga}, makes this blazar an interesting case study.  At the highest observed energies, this source is screened from the observer by EBL with an optical depth $\tau>5$~\cite{Furniss:2013kv}, even for the lowest level of EBL.  It is highly unlikely that any primary gamma rays, originating at the source, would be observed from this object, and, therefore, it is of interest to compare the observed spectrum with theoretical predictions of secondary gamma rays. 

\section{PKS~1424+240 and secondary gamma rays}

The spectrum of secondary gamma rays is remarkably insensitive to the model parameters~\cite{Essey:2009ju,Essey:2010er}, except for the overall normalization determined by the source power in $E_p>10^{17}$~eV cosmic rays.  The latter parameter is uncertain, although spectral fits to other sources were obtained for values consistent with theoretical models of cosmic rays~\cite{Essey:2009ju,Essey:2010er,Razzaque:2011jc}.  There are many models of EBL derived using different approaches~\cite{Stecker:2005qs,Franceschini:2008tp,Finke:2009xi,Stecker:2012ta,Inoue:2012bk}. 
For a number of sources, both ``high'' and ``low'' EBL models fit the data well, 
with no statistically significant difference in the goodness of fit~\cite{Essey:2010er}.  However, for more distant sources, the data may be more sensitive to a model of EBL. Furthermore, 
the shape of the secondary spectrum does depend on the redshift.   Only the lower bound exists on the redshift of PKS~1424+240.  Secondary gamma rays of TeV energies can be seen from sources beyond $z=1$~\cite{Aharonian:2012fu}.  We will, therefore, consider two different redshifts consistent with the lower bound.

The dependence of the secondary spectrum on the redshift of the source~\cite{Essey:2011wv} can be easily understood. The electromagnetic showers induced by the interactions continue to soften the spectrum until the gamma rays produced have an energy such that the optical depth $\tau \lesssim 1$.  The optical depth grows with distance, and it also grows with energy. 
Thus the low-energy peak of the secondary spectrum occurs at a lower energy for more distant sources, leading to a softer spectral shape. For two different redshifts, the spectral shapes at high energies are similar, as long as the condition $\tau \gg 1$ holds true for both redshifts.

We will show that some combinations of redshift and EBL give a good fit to the VHE data.  In particular, we compare with the data our theoretical predictions for secondary gamma rays assuming that (i) the source is at  $z=0.6$ and EBL spectrum based on Ref.~\cite{Stecker:2012ta}; (ii) the source is at $z=1.0$ and $z=1.3$ for a lower level of EBL based on Ref.~\cite{Franceschini:2008tp}. Each of these possibilities produces a substantially better fit to the data than a primary gamma ray spectrum obtained in the absence of cosmic rays. This is remarkable because the shape of the high-energy spectrum is quite robust and independent of any model parameters.  This successful fit to the data provides additional credence to the idea that distant gamma-ray sources are dominated by a hard, secondary spectrum. However, the spectral shape and fit to experimental data alone cannot distinguish between the different models of EBL and different redshifts.  One can, however, use flaring and variability data to infer further information on the transition from primary to secondary spectrum and to gain information about both the EBL and the redshift of the source.

While the shape of the spectrum is robust, its normalization depends on the source luminosity in protons $L_{\rm p}$, which is not known.  In what follows, we obtain the best fit for isotropic equivalent source power in protons $L_{\rm p, iso} = (1-10)\times 10^{49}$~erg/s.  The true intrinsic luminosity can be smaller by several orders of magnitude, depending on the jet opening angle $\theta$. 
In particular, for $\theta = 3^\circ $, the required intrinsic luminosity is comparable to the Eddington luminosity of a black hole with a mass $M \sim  (1-8)\times 10^8 M_\odot$.  Of course, 
only a fraction of the jet energy is transferred to high-energy particles.  For a non-spherical accretion with a jet, the Eddington luminosity is not necessarily a limiting factor, and 
there is  growing evidence of super-Eddington luminosities in relativistic outflows in GRBs and in very powerful blazars~\cite{Ghisellini}.  
The required luminosity is consistent with general principles of acceleration, as long as most of the accretion energy is converted efficiently to the kinetic energy of the jet, rather than to thermal radiation of the accretion flow~\cite{Aharonian:2012fu}. 

Some assumptions must be made about the strengths of intergalactic magnetic fields (IGMFs) along the line of sight.  If IGMFs are greater than $\sim 3\times 10^{-14}$~G, they can cause sufficient deflections of protons to diminish a contribution of secondary gamma rays.  For very small magnetic fields, below $10^{-17}$~G, the low-energy part of the spectrum (which is not as robust as the TeV part) may exceed the observations of Fermi~\cite{Essey:2010nd}.  Therefore, we choose to fit the data for IGMFs of the order of $10^{-15}$~G, in the middle of the allowed range inferred from observations~\cite{Essey:2010nd}.

Blazars are known to be variable with the variability correlated across a wide range of energies~\cite{Ulrich:1997}. Secondary gamma rays, on the other hand, would have this variability washed out~\cite{Prosekin:2012ne}, and no variability should be seen in the secondary component on the timescale of the VERITAS observations. Thus it is reasonable to consider the possibility that the observation at lower energies was due to a flaring state, which should typically have a timescale of the order of days to months~\cite{Ulrich:1997}.  The flaring is further supported by Swift data for the VERITAS  observation period~\cite{Acciari:2010,Williams}. One can expect further observations to detect a lower flux state (unless low energy data again points to a flaring state), which should be clearly evident at energies where $\tau \ll 1$. However, the secondary component should not correlate with this lower energy state and would remain roughly constant.

\section{Results}
\begin{figure}[!ht]
\begin{center}
\includegraphics[width=0.95\linewidth]{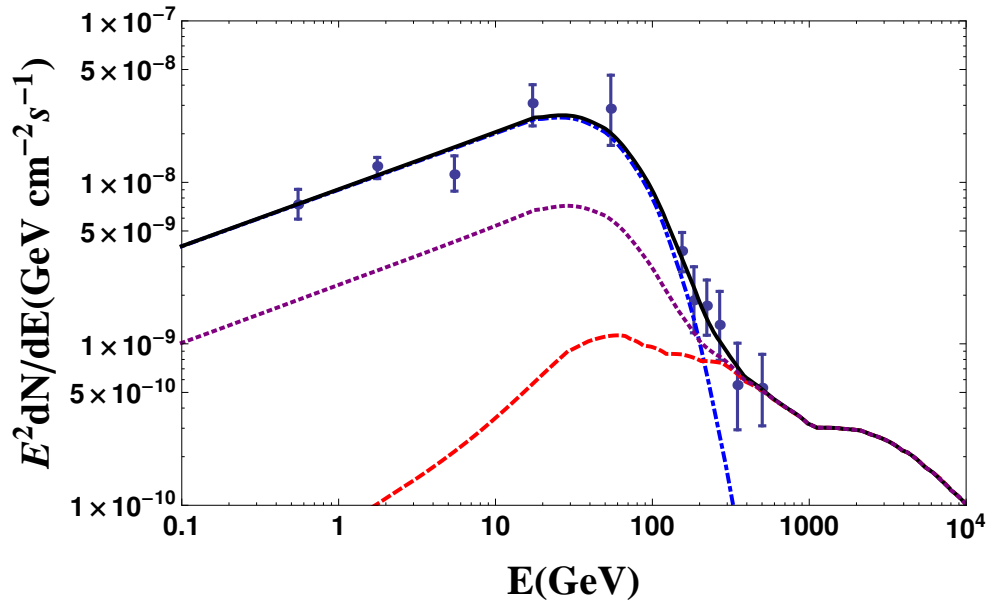}
\caption{Predicted spectrum of PKS 1424+240 (solid black line), assuming $z=0.6$, the lowest redshift consistent with the lower bound~\cite{Furniss:2013kv}, for EBL model of Ref.~\cite{Stecker:2005qs}.  The blue dot-dashed line is primary spectrum (spectral index $\gamma = 1.65$); 
red dashed line is secondary gamma rays for the mean IGMF of $B=10^{-15}$~G with a correlation length of 1~Mpc; black solid line is the combined 
spectrum of primary and secondary gamma rays.  Also shown are the VERITAS and Fermi data points. The purple line represents the total spectrum 
with primary signal suppressed by a factor of 4 (low state).}
\label{fig:stecker:z06}
\end{center}
\end{figure}

\begin{figure}[!ht]
\begin{center}
\includegraphics[width=0.95\linewidth]{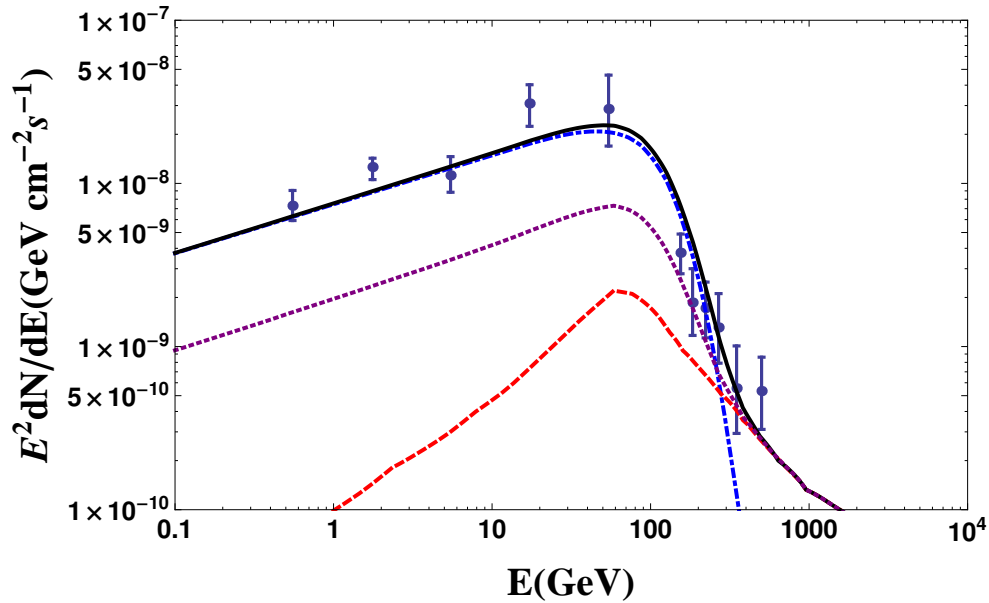}
\caption{Same as Fig.~\ref{fig:stecker:z06}, but assuming redshift $z=1.0$ and using EBL model of Ref.~\cite{Franceschini:2008tp}.}
\label{fig:franc:z10}
\end{center}
\end{figure}

\begin{figure*}[!ht]
\begin{center}
\includegraphics[width=0.95\linewidth]{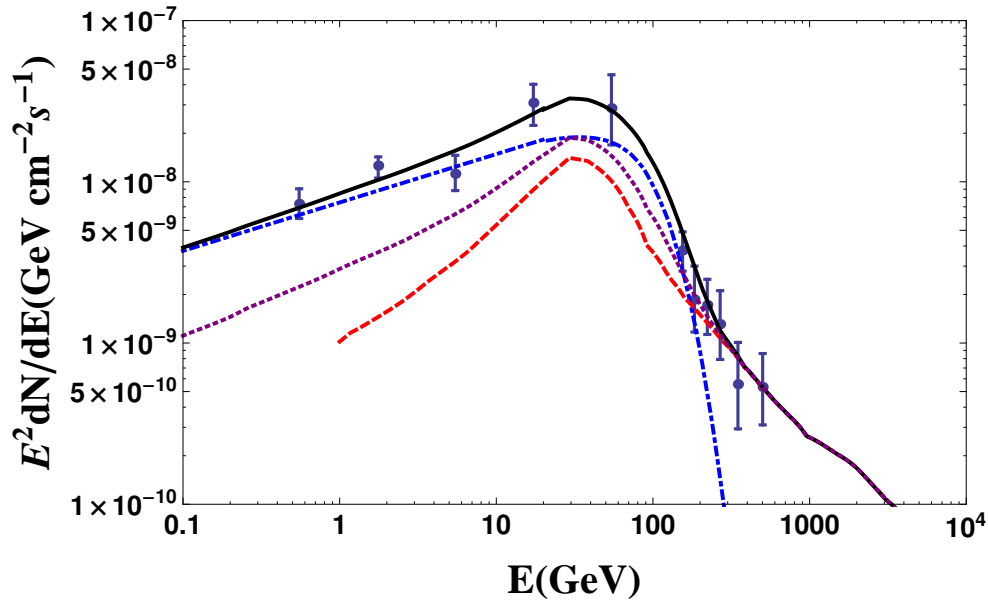}
\caption{Same as Fig.~\ref{fig:stecker:z06}, but assuming redshift $z=1.3$, using EBL model of Ref.~\cite{Franceschini:2008tp}, and 
assuming that primary gamma rays are emitted with a spectral index $\gamma = 1.7$.}
\label{fig:franc:z13}
\end{center}
\end{figure*}

In Figs.~\ref{fig:stecker:z06}--\ref{fig:franc:z13} we illustrate the effect of flaring and show the expected spectra for both a high state and a low state for two different redshift--EBL combinations.
The data are available from VERITAS~\cite{Acciari:2010,Williams} and MAGIC~\cite{Aleksic:2014tga}.  These data are not contemporaneous, and there can be differences in systematic and statistical errors between the two sets of data.  In Figs.~\ref{fig:stecker:z06}--\ref{fig:franc:z13} we show the data from VERITAS, which were accumulated over a longer period of observation and have smaller error bars.  Our best-fit curves that agree with VERITAS data fit the MAGIC data as well.  
 
In each case the flux of the primary component is decreased by a factor 4 (which is reasonable for illustrative purposes and is consistent with observations~\cite{Acciari:2010,Williams,Aleksic:2014tga}), while the secondary component remains constant. Although the flaring spectral data revealed little difference between the two scenarios, a comparison of the high and low states shows some marked differences. Firstly, the spectral shape of the higher redshift source's low state  is significantly softer than the lower redshift source. Secondly, the ratio of integral flux of the high and low states once again differs significantly for the two scenarios.

\begin{figure*}[!ht]
\begin{center}
\includegraphics[width=0.95\linewidth]{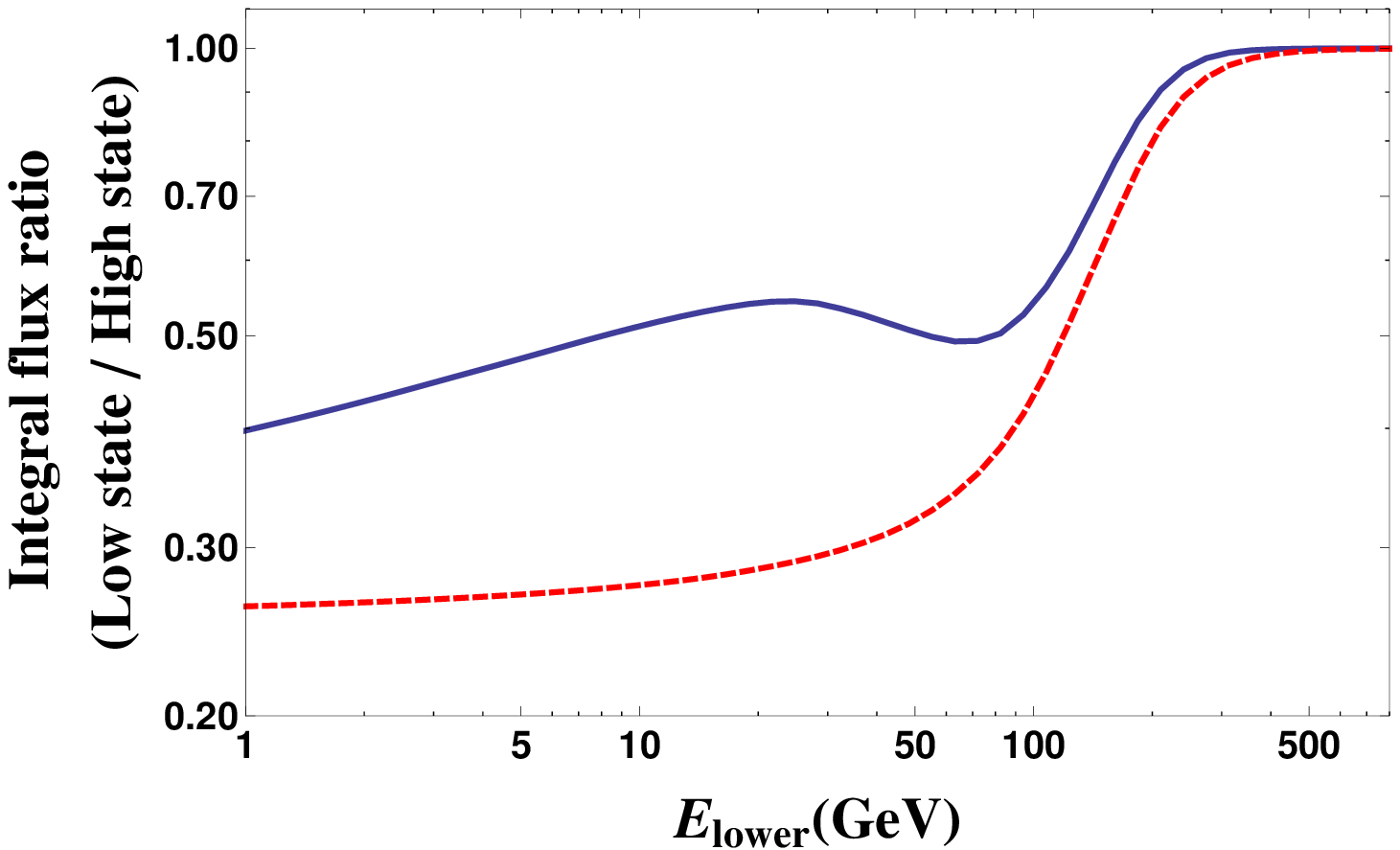}
\caption{Ratio of integral flux of flaring state to low state, where the integral flux is integrated from $E_{\rm lower}$ to 1~TeV. In each the flux contains both a primary and secondary gamma ray component. The primary component is suppressed by a factor of 4 for the low state whereas the secondary component remains constant. The blue solid line shows the results for a source at $z=1.3$ and EBL model based on Ref.~\cite{Franceschini:2008tp} and the red dashed line shows the results for a source at $z=0.6$ and EBL model based on Ref.~\cite{Stecker:2005qs}  }
\label{fig:fluxratio}
\end{center}
\end{figure*}

Figure~\ref{fig:fluxratio} shows the ratio of integral fluxes for the high redshift and the lower redshift scenarios. For the more distant source we expect the secondary component to become a significant component of the overall signal starting from a lower energy. This is clearly seen in the plot and shows that the integral flux, even in the $1-50$ GeV regime, will differ quite significantly from the lower redshift source. Our choice of a low state suppressed by a factor of 4 shows a difference in integral flux ratios of roughly 2 for the energy range $1-50$ GeV.

Thus a measurement of integral flux taken during different states of flux, flaring and low, can reveal important information about the redshift of the source and will lead to an upper limit on the redshift of the source. The integral flux ratios are affected to a lesser degree by the choice of EBL model. Thus it is unlikely that a single source will set a strong limit on the EBL. However, the measurement of multiple sources' flaring and low states will provide valuable information about the EBL, especially for sources with known redshifts.

\section{Conclusions}
To summarize, the inclusion of a secondary gamma ray component, produced by line-of-sight cosmic ray interactions, provides a good fit for the most distant TeV gamma ray source PKS~1424+240 in a broad range of redshifts, from the observational lower limit $z>0.6$~\cite{Furniss:2013kv} to redshifts as high as $z=1.3$.  Future observations of PKS 1424+240 can be used to set more restrictive limits on its redshift, especially if time variability is observed for the lower-energy part of the spectrum.   The combination of spectral and integral flux information for flaring and low states can be used to set limits on the EBL using the future observations of this and other distant blazars.  While at present PKS~1424+240 is the most EBL-obscured known  source, one expects to detect additional TeV sources at redshifts $z\sim 1$~\cite{Aharonian:2012fu}, especially with the advent of the 
Cherenkov Telescope Array (CTA) capable of an unbiased all-sky survey~\cite{Inoue:2013vpa}.  A combined analysis of spectral and temporal information,  similar to our analysis of PKS~1424+240, will be a powerful probe of EBL, when applied to the large dataset expected in the near future~\cite{Aharonian:2012fu,Inoue:2013vpa}.  

\subsubsection*{Acknowledgements:} 

This work was supported by the U.S. Department of Energy Grant DE-SC0009937 and by the World Premier International Research Center Initiative (WPI Initiative), MEXT, Japan.






\end{document}